\documentclass{jpsj3}

\makeatletter
\newcommand\newblock{\hskip .11em\@plus.33em\@minus.07em}
\makeatother

\usepackage{txfonts}
\usepackage{authblk}
\usepackage{url}

\usepackage{graphicx}
\graphicspath{ {images/} }

\title{Comparing the Effects of Boltzmann Machines as Associative memory in Generative Adversarial Networks between Classical and Quantum Samplings}
\author[1,2]{Mitsuru Urushibata}
\author[1,3,4]{Masayuki Ohzeki}
\author[1]{Kazuyuki Tanaka}
\affil[1]{Graduate School of Information Sciences, Tohoku University, Sendai 980-8579, Japan}
\affil[2]{Crosstab Inc., Kawasaki 211-0068, Japan}
\affil[3]{Sigma-i Co., Ltd., Tokyo 108-0075, Japan}
\affil[4]{Institute of Innovative Research, Tokyo Institute of Technology, Yokohama 226-8503, Japan}

\abst{\qquad We investigate the quantum effect on machine learning (ML) models exemplified by the Generative Adversarial Network (GAN), which is a promising deep learning framework. 
In the general GAN framework, the generator maps uniform noise to a fake image. 
In this study, we utilize the Associative Adversarial Network (AAN), which consists of a standard GAN and an associative memory. Moreover, we set a Boltzmann Machine (BM), which is an undirected graphical model that learns low-dimensional features extracted from a discriminator, as the memory.  
Owing to the difficulty of calculating the BM's log-likelihood gradient, it is necessary to approximate it by using the sample mean obtained from the BM, which has tentative parameters. 
To calculate the sample mean, a Markov Chain Monte Carlo (MCMC) method is often used.
In a previous study, this was performed using a quantum annealing device, and the performance of the ``Quantum" AAN was compared with that of the standard GAN.
However, its better performance than the standard GAN is not well understood. 
In this study, we introduce two methods to draw samples: classical sampling via MCMC and quantum sampling via quantum Monte Carlo (QMC) simulation, which is quantum simulation on the classical computer. 
Then, we compare these methods to investigate whether quantum sampling is advantageous. 
Specifically, calculating the discriminator loss, generator loss, inception score, and Fréchet inception distance, we discuss the possibility of AAN. 
We show that the AANs trained by both MCMC and QMC are more stable during training and produce more varied images than the standard GANs. 
However, the results indicate no difference in sampling by QMC simulation compared with that by MCMC.}

\inst{}
\begin{document}
\maketitle

\section{Introduction}
In recent years, researchers have proposed various types of deep learning methods and have applied them in diverse areas. For example, the deep convolutional neural network (CNN)\cite{krizhevsky2012imagenet} is used for image recognition, and the recurrent neural network (RNN)\cite{soltau2016neural} for speech recognition.

The Generative Adversarial Network (GAN)\cite{goodfellow2014generative,wang2019generative} is a promising deep learning framework that comprises a generative model consisting of two multilayer deep neural networks---a discriminative model \textit{D} and a generative model \textit{G}.
The discriminative model \textit{D} learns input images to distinguish between real and fake images, which are provided by \textit{G}. 
On the other hand, the generative model \textit{G} providing fake images is trained to deceive \textit{D} into judging fake images as real. 
GAN generates images that seem very real and does not need training labels. 
Consequently, it is widely used in real-world applications. 

However, several studies \cite{salimans2016improved,metz2016unrolled}, have reported problems with training GANs owing to the two networks learning simultaneously.
For example, it is difficult for both the generator's and discriminator's loss function to converge to an equilibrium point, the learning process results in mode collapse, where \textit{G} produces a small variety of outputs. 
To avoid the mode collapse, GANs such as Wasserstein GAN\cite{arjovsky2017wasserstein} and unrolled GAN\cite{metz2016unrolled} have consequently been introduced. In addition, feature matching and minibatch discrimination \cite{salimans2016improved} have been proposed.

Arici and Celikyilmaz\cite{Arici2016} proposed a completely different approach that employs a third network to connect \textit{D} and \textit{G}, called associative memory.
The overall framework is called the Associative Adversarial Network (AAN). 
To prevent \textit{D} from learning earlier than \textit{G}, the associative memory learns low-dimensional representations of an image extracted by \textit{D} and generates samples from the memory. 
Then, the samples are input into \textit{G} to produce fake images.

A Boltzmann Machine (BM) is a stochastic neural network that has a probability distribution given by an energy function. 
A Restricted BM (RBM) consists of hidden nodes and visible nodes, with nodes of the same type being independent of each other. In general, it is difficult to calculate the gradient descent of the negative log likelihood of the BM because the gradient has two terms, one of which is intractable. Thus, the term is approximated by the sample mean. Particularly for RBMs, Contrastive Divergence (CD)\cite{hinton2002training} is a useful and easy sampling method owing to the structure of RBM.
Arici and Celikyilmaz showed that AAN using RBM as the associative memory produces samples that alleviate the learning task.

In several studies\cite{anschuetz2019near,Wilson2021}, quantum sampling methods were used when training the BM acting as the AAN's memory. 
One such approach\cite{Wilson2021} applied a quantum annealer (D-Wave2000Q) to draw samples from a Boltzmann-like distribution instead of the CD method for RBM. The approach is applicable to not only a restricted network topology but also various other networks, such as complete and sparse networks. 
Quantum annealing was originally proposed for solving optimization problems via the quantum tunneling effect \cite{Kadowaki1998,Ray1989,Das2005,Das2008,ohzeki2011quantum}.
However, its implementation in practical devices is affected by the thermal heat bath and results in an approximate Gibbs–Boltzmann sampler for generating the equilibrium state\cite{amin2015searching}.
It has been shown that the quantum AAN slightly outperforms the classical AAN when using the Gibbs sampler to draw samples from the BM for an Modified National Institute of Standards and Technology (MNIST)\cite{dmnist} dataset---a large database of handwritten digits. 
Another approach using classical quantum sampling\cite{anschuetz2019near} has also shown that a quantum AAN works better than a classical AAN on the MNIST and Canadian Institute For Advanced Research (cifar)10 datasets (the cifar10 dataset consists of colored images). 
These results indicate that quantum sampling has a slightly positive effect on the GAN.    

The application of the quantum annealer to combinatorial optimization problems has been widely studied\cite{rosenberg2016solving,venturelli2015quantum,neukart2017traffic,Kumar2017,Takahashi2018,Ohzeki2018NOLTA,Ohzeki2019, nishimura2019item, Yonaga2020solving, ide2020maximum, Arai2021,Sato2021, Koshikawa2021}.
The quantum effect on the degenerate ground state has also been investigated \cite{Yamamoto2020, maruyama2021graph}. 
Several groups have reported sampling applications of quantum annealing to machine learning\cite{amin2018quantum,Kumar2017,adachi2015application,benedetti2016estimation}. 
Related comparative studies have also been performed with benchmark tests for solving optimization problems \cite{Oshiyama2022}.
To investigate the further possibilities of its application, it is important to clarify if there are any quantum effects on AANs. 
In the present study, we assessed whether the quantum effect contributes to the learning of images by AAN through a series of experiments. In this paper, we first introduce basic concepts regarding the GAN framework, including AANs, BM, and quantum BM (Sect. 2). In Sect. 3, we describe the experiments conducted and give an overview of the results obtained. Finally, we analyze the results and present further issues in Sect. 4.       

\section{Preliminaries}
\subsection{Generative Adversarial Network}
A GAN\cite{goodfellow2014generative} consists of two neural networks (a generative model and a discriminative model) and learns the distribution $p_g$ over the given data $\bold{x}$. 
The generative model \textit{G} provides fake images similar to $\bold{x}$, and the discriminative model \textit{D} distinguishes between fake and real images $\bold{x}$. 
To learn $p_{g}$, the weight parameters of \textit{G} are updated to deceive \textit{D} and those of \textit{D} are updated to correctly identify real and fake images. 
Let $G(\bold{z},\boldsymbol{\theta}^{(G)})$ be a function that produces a fake image from input $\bold{z}$ drawn from uniform distribution $p_{\bold{z}}$, and $D(\bold{x},\boldsymbol{\theta}^{(D)})$ be a function that returns one if it is a real image and zero otherwise. 
Letting $p_{data}$ be an empirical data distribution, the objective function of the GAN is defined as

\begin{eqnarray}
\label{GAN}
    V(D,G) = \mathbb{E}_{\bold{x}\sim p_{data}(\bold{x})}[\log D(\bold{x})] + \mathbb{E}_{\bold{z}\sim p_{\bold{z}}(\bold{z})}[\log (1-D(G(\bold{z}))].
\end{eqnarray}
GAN's learning algorithm can be formulated as the min--max game for $\min_{G} \max_{D}V(D,G)$.

\subsection{Boltzmann machine}
A BM is an undirected graph model that has its unit state probability given by an energy function. 
Let $G=(\Omega,E)$ be the graph, the energy function is defined as
\begin{eqnarray}
\label{hamiltonian}
E(\boldsymbol{z};\boldsymbol{\theta}) =  - \sum_{i \in \Omega }b_{i}z_{i}  -\sum_{(i,j)\in E}w_{ij}z_{i}z_{j},
\end{eqnarray}
where $\boldsymbol{\theta} = (\boldsymbol{w},\boldsymbol{b})$, $w_{ij}$ is the connection strength between $z_{i}$ and $z_{j}$, $b_{i}$ is the bias of $z_{i}$, and $z_{i}$ are states of the $i$-th unit, taking values \{-1,1\}. 
The probability distribution of $\boldsymbol{z}$ is defined by
\begin{eqnarray}
P(\boldsymbol{z})= \frac{\exp(-\beta E(\boldsymbol{z};\boldsymbol{\theta}))}{Z(\boldsymbol{\theta})} ,
\end{eqnarray}
where $\beta$ is an inverse temperature, $Z(\boldsymbol{\theta}) = \Sigma_{\boldsymbol{z}} \exp(-\beta E(\boldsymbol{z};\boldsymbol{\theta}))$.
Given the data $(\boldsymbol{z}_{1},\boldsymbol{z}_{2},...,\boldsymbol{z}_{N})$, where $N$ denotes the amount of data, the BM is trained using the maximum likelihood method, with the log likelihood given as

\begin{eqnarray}
\log L(\boldsymbol{\theta})= \sum_{k=1}^{N}\left(\beta \left(\sum_{i\in \Omega}b_{i}z_{i}^{(k)} + \sum_{(i,j) \in E}w_{ij}z_{i}^{(k)}z_{j}^{(k)}\right)\right)-N\log Z(\boldsymbol{\theta}), 
\end{eqnarray}
where $z_{i}^{(k)}$ is the $i$-th node of $\boldsymbol{z_{k}}$. 
The maximization of the log likelihood can be performed using gradient ascent. 
The partial derivatives of the log likelihood are written as
\begin{eqnarray}
\frac{\partial\log L(\boldsymbol{\theta})}{\partial b_{i}} &=&  \beta \left(\sum_{k=1}^{N}z_{i}^{(k)}-N\mathbb{E_{p(\boldsymbol{z})}}[Z_{i}]\right), \\
\frac{\partial\log L(\boldsymbol{\theta})}{\partial w_{ij}} &=&  \beta \left(\sum_{k=1}^{N}z_{i}^{(k)}z_{j}^{(k)}-N\mathbb{E_{p(\boldsymbol{z})}}[Z_{i}Z_{j}]\right).
\end{eqnarray}
For $n$ spins, we must calculate the sum of $2^n$ terms to obtain the expected value of the second terms, but it is intractable. 
Instead, we use the approximation $\mathbb{E_{p(\boldsymbol{z})}}[Z_{i}] \approx \sum_{k=1}^S x_{i}^{(k)}/S$ using samples $(\boldsymbol{x}_{1},\boldsymbol{x}_{2},...,\boldsymbol{x}_{S})$, where $S$ denotes the sample size drawn from the BM parameters before updating.
To do this, we use Gibbs sampling---a Markov chain Monte Carlo (MCMC) algorithm. 
The proposal density of the BM is calculated as
\begin{eqnarray}
p(x_{i}^{'}|\Omega\setminus{\{x_{i}\}}) = \frac{\exp(\beta x_{i}^{'} \lambda)}{\exp(-\beta\lambda) + \exp(\beta\lambda)},
\end{eqnarray}
where $x_{i}^{'}$ is the candidate of the $i$-th node and $\lambda $ is $\beta_{i} + \sum_{\{j|(i,j) \in E\}} w_{ij}x_{j}$. 
Let $r$ be a uniform random number on (0,1]; if $p(x_{i}^{'}=-x_{i}) \geq r$, then the $i$-th node is flipped; this is repeated for the other nodes until convergence. 

\subsection{Quantum Monte Carlo Simulation}
The quantum Monte Carlo (QMC) simulation is a partial simulation of QA device sampling on a classical computer. 
Only some aspects of the QA device are sampled; in particular, low-energy states in the equilibrium state, can be demonstrated by the QMC method \cite{Bando2021}.
The equilibrium state of the quantum annealer is described as the transverse field quantum Ising model. The Hamiltonian of this model is defined as 
\begin{eqnarray}
\hat{H} = \hat{H_{\boldsymbol{\lambda}}}+\hat{H_{\boldsymbol{\perp}}}, 
\end{eqnarray}
where $\hat{H_{\boldsymbol{\lambda}}}$ is derived to map the Hamiltonian of the classical BM Eq. (\ref{hamiltonian}) to the Ising model for the quantum system as
\begin{eqnarray}
    \label{quantum hamiltonian}
    \hat{H}_{\boldsymbol{\lambda}}(\boldsymbol{\sigma}) = -\sum_{i\in \omega}h_{i}\hat{\sigma_{i}}^{z}-\sum_{(i,j)\in E}J_{ij}\hat{\sigma_{i}}^{z}\hat{\sigma_{j}}^{z}, 
\end{eqnarray}
where $\boldsymbol{\lambda}$ is $(\boldsymbol{h},\boldsymbol{J})$ and $\hat{\sigma_{i}}^{z}$ is the $z$ component of the Pauli matrices operating on the $i$-th qubit. 
The transverse field term $\hat{H_{\perp}}$ is defined as
\begin{eqnarray}
    \hat{H_{\boldsymbol{\perp}}}  =-\Gamma(t)\sum_{i \in \omega}\hat{\sigma_{i}}^{x}, 
\end{eqnarray}
where $\Gamma(t)$ is the magnitude of the transverse field. The probability distribution of the spin states can be written as
\begin{eqnarray}
    \label{dist}
    \hat{\rho} = \frac{e^{-\beta(\hat{H_{\boldsymbol{\lambda}}}+\hat{H_{\boldsymbol{\perp}}})}}{Z}, 
\end{eqnarray}
where $Z=\mathrm{tr}(e^{-\beta(\hat{H_{\boldsymbol{\lambda}}}+\hat{H_{\boldsymbol{\perp}}})})$.
With the Suzuki--Trotter expansion, the numerator of Eq.~(\ref{dist}) is written as $e^{-\beta(\hat{H_{\boldsymbol{\lambda}}}+\hat{H_{\boldsymbol{\perp}}})} \thickapprox (e^{-\frac{\beta}{M}\hat{H_{\boldsymbol{\lambda}}}}e^{-\frac{\beta}{M}\hat{H_{\boldsymbol{\perp}}}})^{M}$. For a large volume of $M$, the transverse field quantum  $L$-dimensional Ising model is mapped to the classical $L$+1-dimensional Ising model with the Hamiltonian as 
\begin{eqnarray}
    H(\boldsymbol{\sigma})=-\frac{1}{M}\sum_{m=1}^{M}\sum_{i \in \omega} h_{i}
    \sigma_{i,m}-\frac{1}{M}\sum_{m=1}^{M}\sum_{(i,j
)\in E} J_{ij}\sigma_{i,m}\sigma_{j,m}-\frac{1}{2 \beta}\log \coth\left( \frac{\beta \Gamma}{M}\right) \sum_{m=1}^{M}\sum_{i \in \omega} \sigma_{i,m}\sigma_{i,m+1}.
\end{eqnarray}
When $M\gg1$, $\sigma_{i,m}$ denotes the $i$-th node of the $m$-th Trotter layer. 
The QMC draws samples from the state distribution defined as $\exp(-\beta H(\boldsymbol{\sigma}))/Z$, where $Z=\sum_{\sigma_{1}\sigma_{2}...\sigma_{M}} \exp({-\beta H(\boldsymbol{\sigma})}) $. 
For a candidate state, $\boldsymbol{\sigma^{'}}$, setting the acceptance probability as $r = \min \left( 1,H(\boldsymbol{\sigma^{'}})/H(\boldsymbol{\sigma})\right)$, we can sample from the Ising model easily without calculating $Z$ in the same manner as the Metropolis--Hastings algorithm.
By using the QMC simulation, we investigate the equilibrium state even with the transverse field.
When we set $\beta$ to be a large value, the equilibrium state converges to the ground state (the lowest energy state).
In the BM learning, $\beta$ is absorbed into the biases and interactions.
In addition, the output from the D-Wave quantum annealer approximately obeys the Gibbs--Boltzmann distribution.
Note that QMC is, in essence, a method to investigate the equilibrium state of the quantum many-body system.
QMC does not necessarily reproduce the nonequilibrium behavior in the D-Wave quantum annealer \cite{Bando2021}.
The BM utilizes the equilibrium distribution to sample the spin configuration.
In this sense, the performances of the AAN and quantum AAN are purely compared by the equilibrium states of both the classical and quantum equilibrium states.
In a strict sense, the last stage of the quantum annealing is similar to that of classical annealing without any quantum or thermal effects.
Thus, we do not expect any differences to appear in both of the methods.
Therefore, we theorize that the performance shown in the previous study is purely the Boltzmann sampling in the random input generator.
Unfortunately, in the previous study\cite{Arici2016} did not compare their method with the standard GAN and the AAN with classical Monte Carlo sampling.
Our aim is to investigate the QMC sampling as well as the classical sampling on the standard GAN to clarify the advantage of quantum AAN, as proposed in the previous study.
Below, we outline several experiments conducted to investigate any differences among various combinations used to generate the data using the standard GAN, the classical AAN, and the quantum AAN by QMC.
The difference between classical Monte Carlo and QMC simulations implies that nontrivial quantum effects exist in sampling even in the equilibrium state.
No difference suggests that the quantum effect comes from the difference between the equilibrium state and the resulting state from the quantum annealer, if it exists.

\subsection{Associative Adversarial Network}
As stated above, Arici and Celikyilmaz\cite{Arici2016} introduced a new GAN framework called Associative Adversarial Network (AAN) using memory that learns the low-dimensional image features extracted by $D$. 
The AAN framework uses the samples drawn from the associative memory as input to $G$,
instead of using the uniform noise on $[0,1]$ as in the standard GAN. 
BM has been used as an associative memory by Arici and Celikyilmaz and several other groups since then \cite{Arici2016,Wilson2021,anschuetz2019near}. Similar to Eq.~(\ref{GAN}), $D$, $G$, and the associative memory are trained by optimizing the objective functions defined as
\begin{eqnarray}
    V(G,D,\hat{\rho})=\mathbb{E}_{x\sim p_{data}(x)}[\log D(x)]+\mathbb{E}_{f \sim \hat{\rho}(f)}[\log (1-D(G(f)))]+\mathbb{E}_{f \sim \rho(f)}[\log \hat{\rho}(f)],
\end{eqnarray}
where $\rho$ is the true distribution of low-dimensional image features ($f$) extracted by $D$, and $\hat{\rho}$ is the model distribution used as an associative memory.    
We find the solutions to the min-max-max game for $\min_{G} \max_{\hat{\rho}} \max_{D} V(G,D,\hat{\rho})$. Figure \ref{fig:strcture} shows an image of the implemented structure.

In the experiments reported below, we trained AANs on a classical computer using both QMC and MCMC methods and compared their performances.

\section{Experiments}
We perform experiments to investigate whether the effect of quantum sampling has more impact on GAN training than the work of associative memory. 
We trained three models on the MNIST dataset:  standard GAN, AAN using the classical sampling via MCMC, and AAN using quantum sampling via QMC.
The standard GAN is almost identical to the original GAN\cite{goodfellow2014generative} and uses uniform noise $z$ as input to $G$, but has a few structural differences (to facilitate normal training under the same conditions as the other AANs). 
We subsequently compared the stability of the convergence of learning and the quality of the generated images.
\subsection{Settings}
\subsubsection{Networks}
\begin{figure}[htbp]
    \centering
    \includegraphics[clip, width=15.0cm]{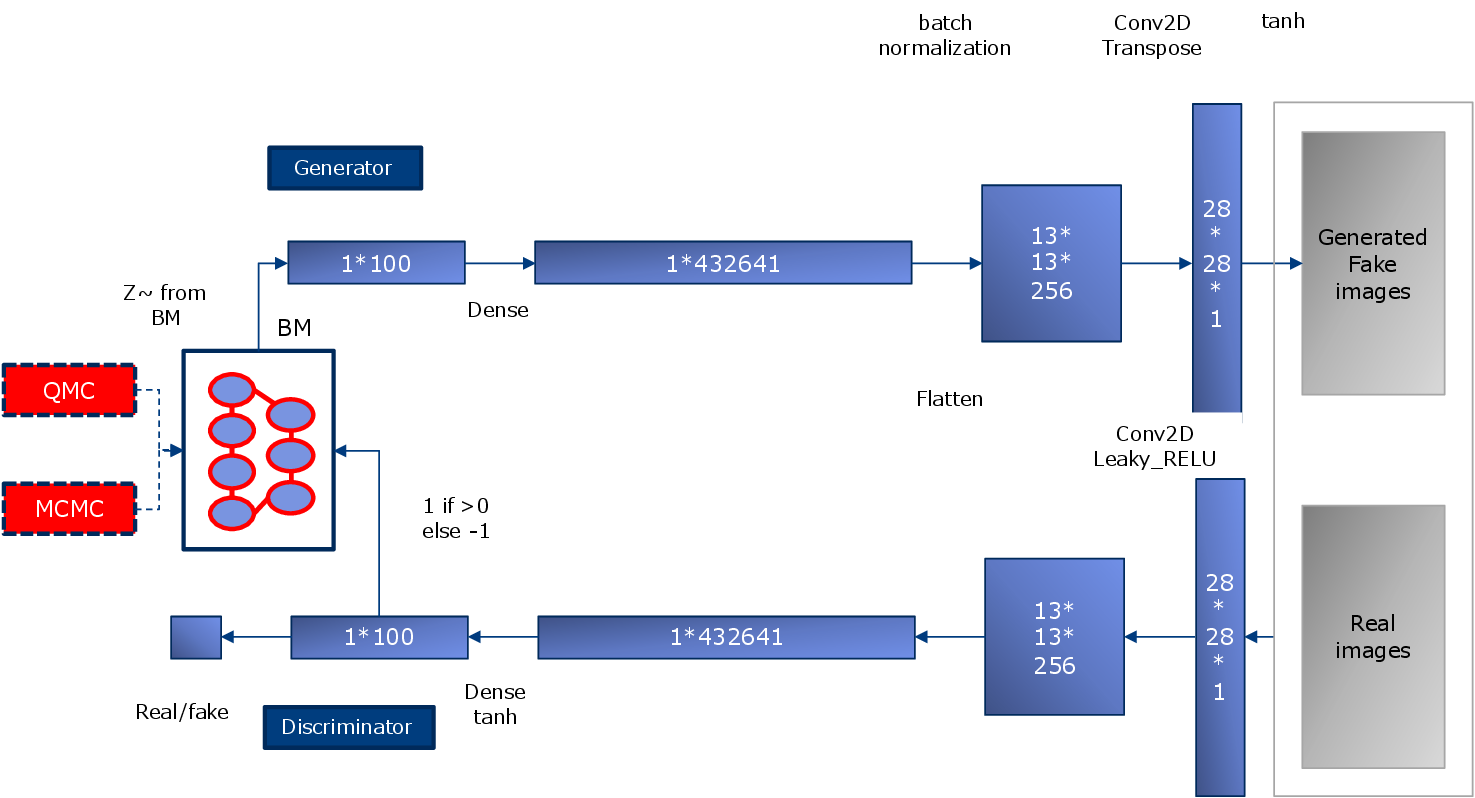}
    \caption{(Color) Network structure in our experiment. The top half of the figure is the generator, and the bottom is the discriminator. Each network has a dense layer and a Conv2D layer. These networks are tied together through a BM.}
    \label{fig:strcture}
\end{figure}
\begin{figure}[htbp]
    \centering
    \includegraphics[clip, width=5.5cm]{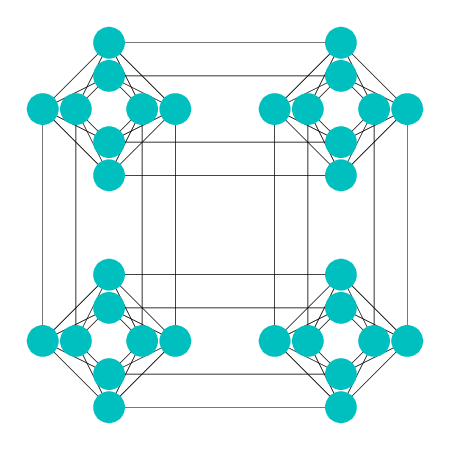}
    \caption{(Color) Structure of sparse network topology with four unit cells.}
    \label{fig:sparse}
\end{figure}

The generator consists of a dense layer and a transposed convolutional layer, which is an output layer. 
As the value of a pixel in the image data is in the range from $-1$ to $1$, the output activation function is tanh. \textit{D} has almost the same layers, and the output activation function is sigmoid. Figure \ref{fig:strcture} provides additional details. Associative memory is located between \textit{G} and \textit{D}. In this experiment, we used three network topologies: complete, sparse, and bipart. The sparse network topology is called a chimera graph\cite{dwavesystem}. The structure has a set of unit cells consisting of two sets of four qubits, as shown in Fig. \ref{fig:sparse}. Each qubit in one set of four qubits is connected to all the qubits in another set. The qubits in the same set are not connected to each other. One set of four qubits in a unit cell is connected to a set of qubits in another horizontally adjacent unit cell, and the other is connected to a vertically adjacent unit cell.  

\subsubsection{Learning}
We used the optimizer Adam while training both \textit{D} and \textit{G}, with the former having a learning rate of 0.0002 and the latter having a learning rate of 0.0004. The batch size was 128 and the number of iterations was 5000. In addition, label smoothing was applied to avoid overfitting as follows: $\mathrm{True\:label} = (1-\mathrm{smoothing\:parameter}) + 0.5(\mathrm{smoothing\:parameter}), \mathrm{False\:label}  = 1-\mathrm{True\:label}$; the parameter was set to 0.2. 
Furthermore, in this experiment, we trained three types of BM, each with a different network topology, using Gibbs sampling, which is an MCMC algorithm and QMC, respectively. These hyperparameters are summarized in Table 1. Because the nodes of BM take a value of either $-1$ or $1$, these were transformed to continuous values in the same way as in previous research\cite{Wilson2021}, as follows. Let $z_{i}$ be the ith node of the BM and $r$ be a random number drawn uniformly on [$-1$,$1$], we set $\alpha = - 4$.
\begin{equation}
f(z_{i}) = 
\begin{cases}
    \displaystyle\frac{\exp(-\alpha(1-r))-\exp(-2\alpha)}{1-\exp(-2\alpha)} & (z_{i}=1) \\
    -1 & (z_{i}=-1)
\end{cases}
\end{equation}
See Tables \ref{tb:hp mcmc} and  \ref{tb:hp qmc} for the other hyperparameters.

\begin{table}[htbp]
    \centering
     \begin{tabular}{c|c|c|c} \hline \hline
    & Complete & Sparse & Bipart \\ \hline \hline
    Learning rate for D & \multicolumn{3}{c}{0.0002} \\ \hline
    Learning rate for G & \multicolumn{3}{c}{0.0004} \\ \hline
    Batch size & \multicolumn{3}{c}{128} \\ \hline
    Number of iterations & \multicolumn{3}{c}{5000} \\ \hline
    Learning rate for BM & 0.0004   & 0.002  & 0.0004 \\ \hline
    Inverse temperature  & 0.15     & 0.25   & 0.20   \\ \hline
    \end{tabular}
    \caption{Hyperparameters used while learning classic AANs.}
    \label{tb:hp mcmc}
\end{table}

\begin{table}[htbp]
    \centering
     \begin{tabular}{c|c|c|c} \hline \hline
    & Complete & Sparse & Bipart \\ \hline \hline
    Learning rate for D & \multicolumn{3}{c}{0.0002} \\ \hline
    Learning rate for G & \multicolumn{3}{c}{0.0004} \\ \hline
    Batch size & \multicolumn{3}{c}{128} \\ \hline
    Number of iterations & \multicolumn{3}{c}{5000} \\ \hline
    Learning rate for BM & 0.0000015   & 0.000004  & 0.000002.5 \\ \hline
    Inverse temperature  & \multicolumn{3}{c}{50}   \\ \hline
    Trotter number  & \multicolumn{3}{c}{32}   \\ \hline
    Initial value of transverse field: $\Gamma(0) $  & \multicolumn{3}{c}{10}   \\ \hline
    \end{tabular}
    \caption{Hyperparameters used while learning quantum AANs. There are two additional parameters [Trotter number and $\Gamma(0)$] compared with the classics.}
    \label{tb:hp qmc}
\end{table}

\subsection{Result}
\begin{figure}[htbp]
    \begin{tabular}{c}

      % 1
      \begin{minipage}{0.33\hsize}
        \begin{center}
          \includegraphics[clip, width=5.0cm]{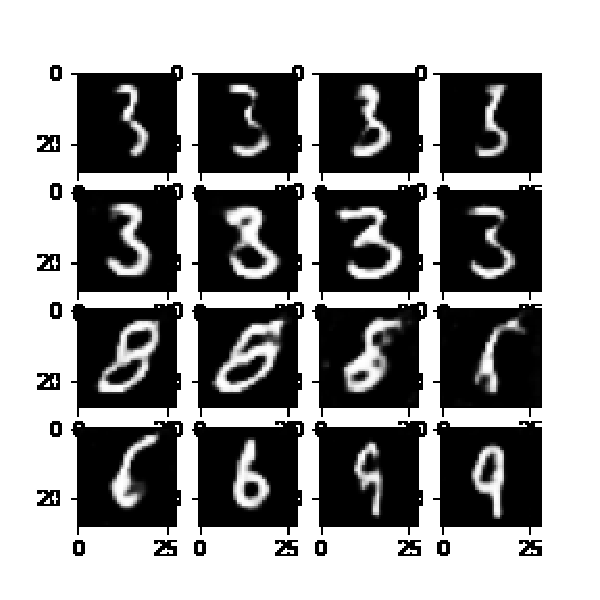}
        \end{center}
      \end{minipage}

      % 2
      \begin{minipage}{0.33\hsize}
        \begin{center}
          \includegraphics[clip, width=5.0cm]{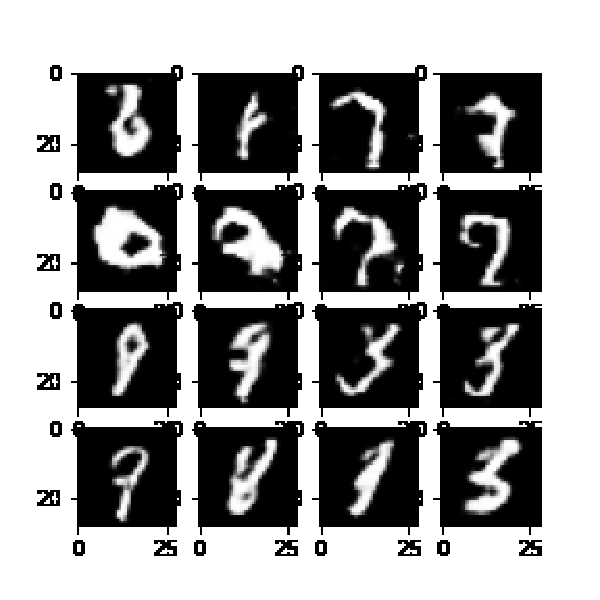}
        \end{center}
      \end{minipage}

      % 3
      \begin{minipage}{0.33\hsize}
        \begin{center}
          \includegraphics[clip, width=5.0cm]{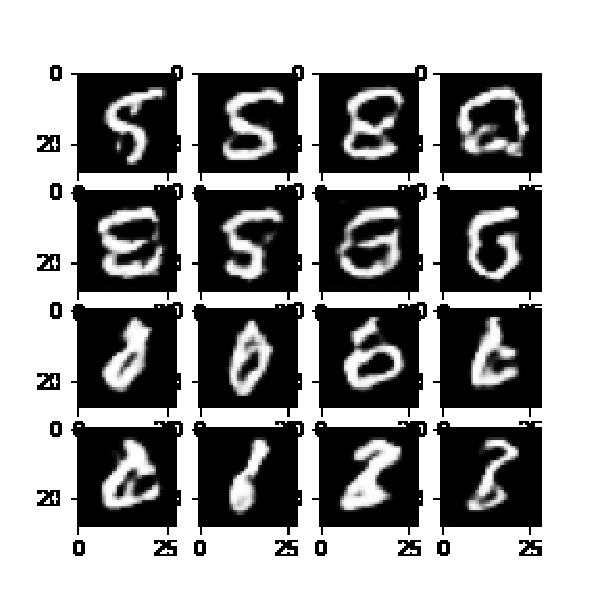}
        \end{center}
      \end{minipage}\\
      
        \begin{minipage}{0.33\hsize}
        \begin{center}
          \includegraphics[clip, width=5.0cm]{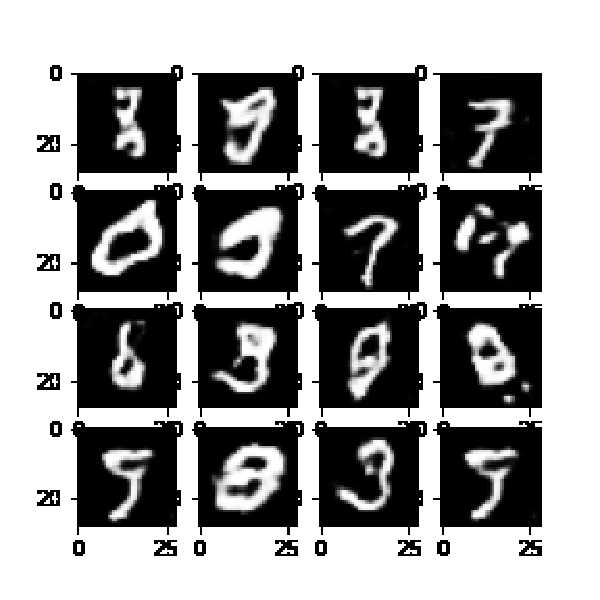}
          \hspace{1.6cm} [a] Complete
        \end{center}
      \end{minipage}
      
            % 2
      \begin{minipage}{0.33\hsize}
        \begin{center}
          \includegraphics[clip, width=5.0cm]{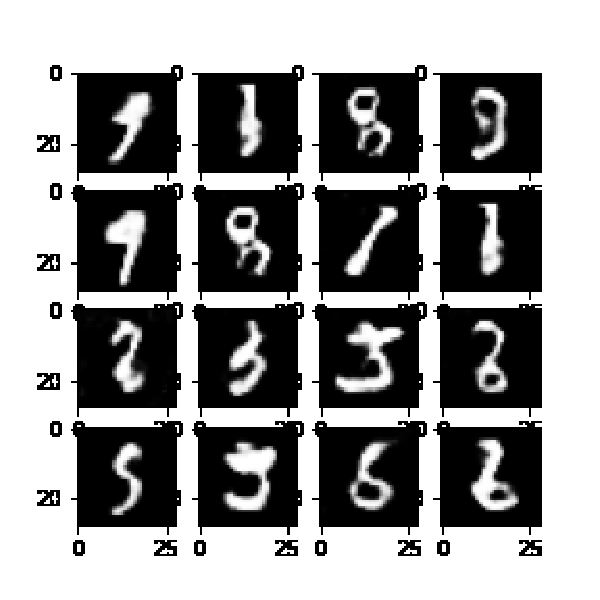}
          \hspace{1.6cm} [b] Sparse
        \end{center}
      \end{minipage}

      % 3
      \begin{minipage}{0.33\hsize}
        \begin{center}
          \includegraphics[clip, width=5.0cm]{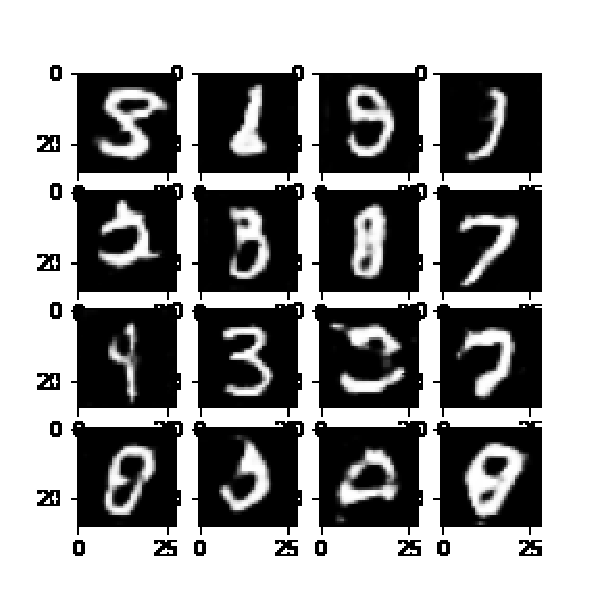}
          \hspace{1.6cm} [c] Bipart
        \end{center}
      \end{minipage}
      
    \end{tabular}
    \caption{(Top row) MNIST image generated using MCMC. (Bottom row) Images generated using QMC . Images (a), (b), and (c) are generated from different BM networks, corresponding to complete, sparse, and bipart, respectively from left to right.}
    \label{fig:images_generated}
\end{figure}
\begin{figure}[htbp]
    \begin{tabular}{c}

      % 1
      \begin{minipage}{0.33\hsize}
        \begin{center}
          \includegraphics[clip, width=5.5cm]{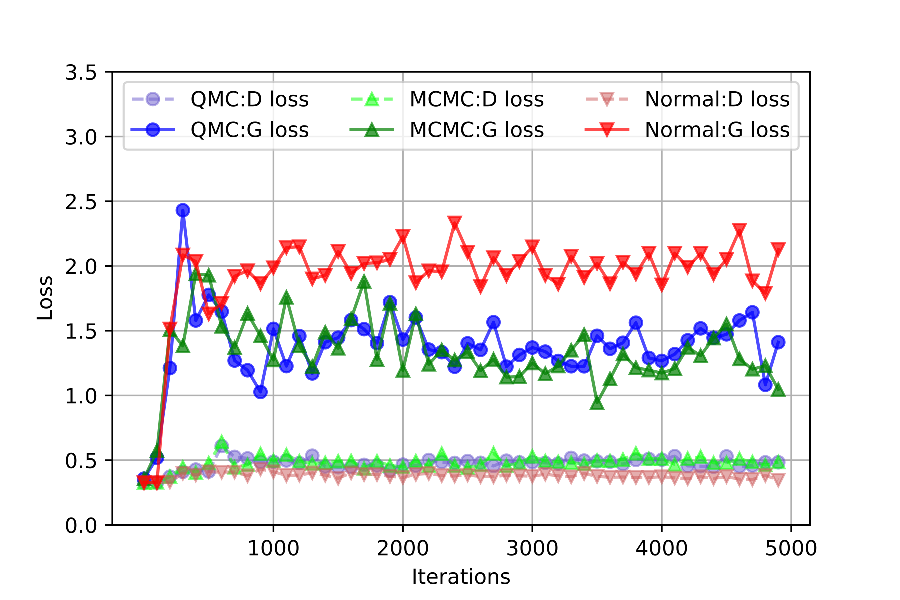}
          \hspace{1.6cm} [a] Complete
        \end{center}
      \end{minipage}

      % 2
      \begin{minipage}{0.33\hsize}
        \begin{center}
          \includegraphics[clip, width=5.5cm]{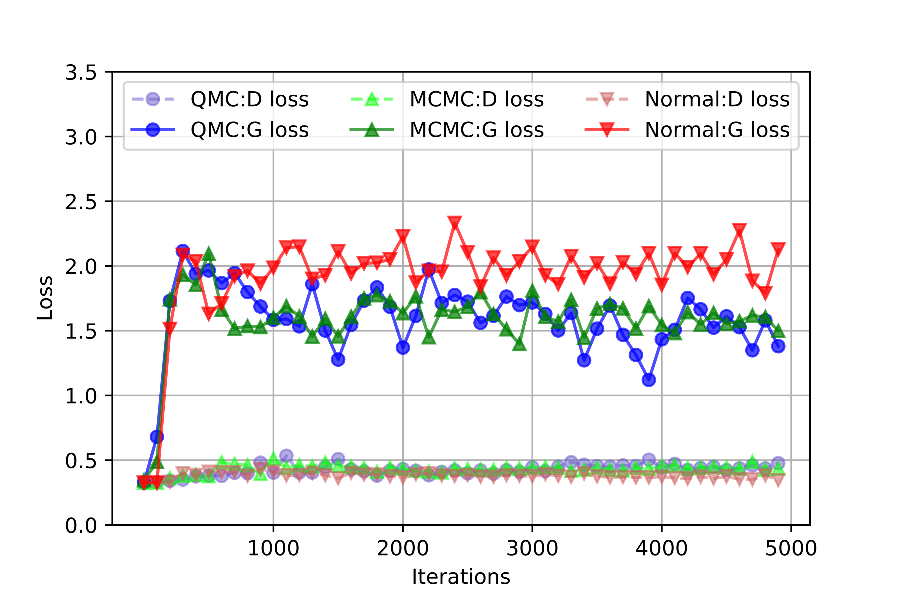}
          \hspace{1.6cm} [b] Sparse
        \end{center}
      \end{minipage}
      % 3
      \begin{minipage}{0.33\hsize}
        \begin{center}
          \includegraphics[clip, width=5.5cm]{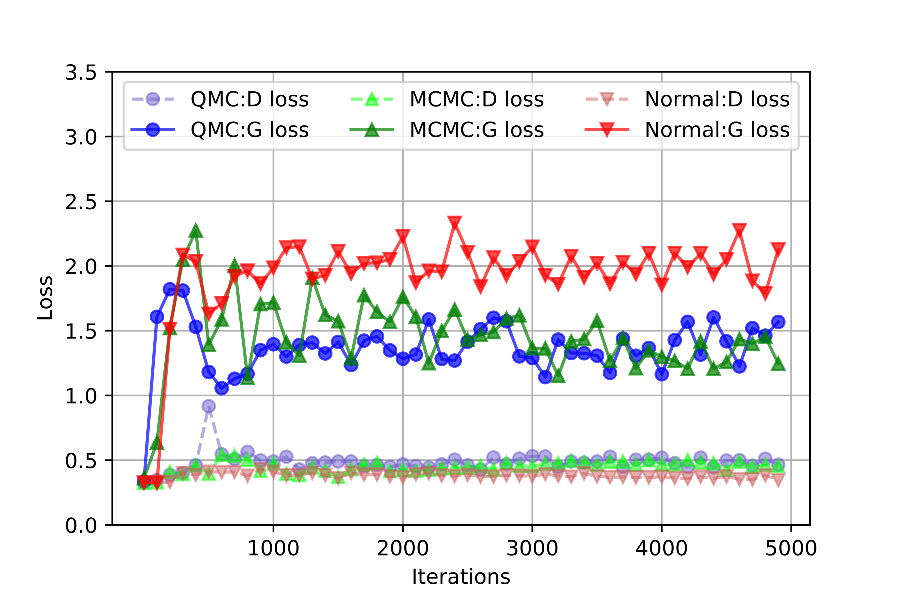}
          \hspace{1.6cm} [c] Bipart
        \end{center}
      \end{minipage}
      \end{tabular}
    \caption{(Color) Losses of discriminator and generator.The horizontal axis shows the number of iterations, and the vertical axis shows the loss of each generator and discriminator (G loss, D loss).}
    \label{fig:loss}
\end{figure}

Figure \ref{fig:images_generated} shows the images generated by each model; they look almost identical. The D loss and G loss are described in Fig. \ref{fig:loss}. The losses are defined as $ -\mathbb{E}_{x\sim p_{data}}[\log(D(x))]/2-\mathbb{E}_{z\sim \hat{\rho}}[\log(1-D(G^*(z)))]/2$, $ -\mathbb{E}_{z\sim \hat{\rho}}[\log(D^*(G(z)))]$, respectively\cite{kodali2017convergence,goodfellow2014generative}. Let $\hat{\rho}$ be the distribution of the BM. The difference between the D loss and G loss is smaller than that of the normal GAN for all three network topologies at each iteration, both using the MCMC and QMC methods.

\begin{figure}[h]
    \centering
    \includegraphics[width=15.0cm]{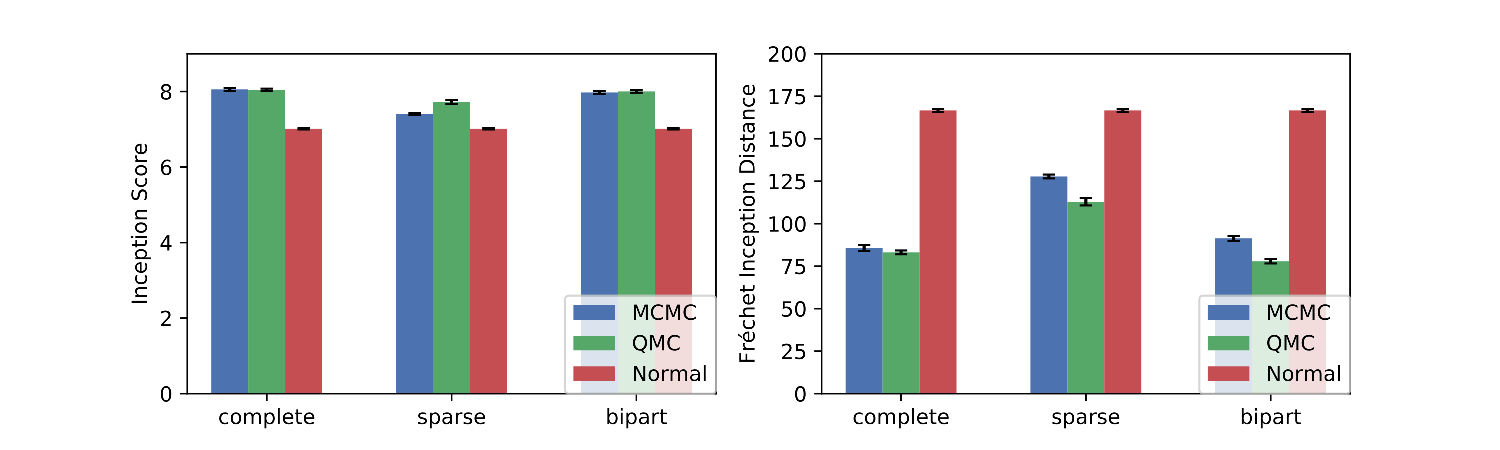}
    \caption{(Color) The left graph shows Inception Score (IS) and the right shows Fréchet Inception Distance (FID). We generated images from the models trained by QMC, MCMC, and Normal GAN, and calculated IS and FID. This was repeated 10 times and the averages of IS and FID are shown in the graph.}
    \label{fig:score}
\end{figure}

The Inception Score (IS) and the Fréchet Inception Distance (FID) are depicted in Fig. \ref{fig:score}. IS, a measure of the quality and diversity of the generated images, is calculated using a pretrained model called the inception v3 network\cite{barratt2018note}, FID\cite{heusel2017gans}, a measure of the distance between the generated images and the training data, is calculated using the same model. However, because the pretrained models used to calculate IS and FID in general have not learned the MNIST dataset, it is impossible to evaluate the images generated from the AANs. Thus, in this experiment, we built a neural network trained to discriminate MNIST handwritten digits, and used it to calculate IS and FID. 

In Fig. \ref{fig:score}, as both the classical AAN and the quantum AAN have higher IS than the normal GAN, it appears that the associative memory is helping to generate a clear and diverse image in the GAN framework. 

As shown above, we find that the AAN has an advantage compared with the standard GAN because we set the BM to generate the random input to the generator G.
However, we do not observe any relevant difference between the classical and QMC simulations.
At the level of generating the equilibrium state by different fluctuations, no quantum merit exists.
It appears that the quantum merit in the previous study on the quantum AAN originated from the nontrivial quantum effect in the quantum annealer.
This is a highly nontrivial problem.

A recent comparative study exhibited subtle differences between the classical simulation of the quantum many-body system and the nonequilibrium behavior of the quantum annealer. 
The remaining possibility as to why quantum AAN has any advantage compared with its classical counterpart stems from the difference in the nonequilibrium behavior.
However, the BM itself is based on the property of the equilibrium state: expectation in equilibrium.
Thus, it is difficult to expect that any advantage of quantum AAN exists with sampling by the quantum annealer.
In this sense, quantum AAN is essentially the same as AAN with classical/quantum MonteCarlo sampling.
Experiments using the sampling of the quantum annealer are more computation-intensive, and will thus be carried out in future studies.

\section{Conclusion and Future work}
This study was inspired by the work of Wilson et al.\cite{Wilson2021}, who trained a classical BM using the D-Wave quantum annealer to generate samples that follow a noisy Gibbs--Boltzmann distribution. In this work, we specifically examined whether quantum effects due to transverse magnetic fields contribute to the learning of GANs by simulating quantum annealing with QMC, except where it depends on properties specific to the D-wave device.

To compare with classical learning methods, we investigated whether there is a difference between classical and quantum samplings for learning AAN. It was reported to be more effective than regular GAN in previous studies\cite{Arici2016}. However, the results of our experiment suggest that there is no advantage to the quantum sampling learning method. Therefore, we conclude that there would be no quantum effect on the performance of AAN at the level of generating the equilibrium state by different fluctuations. 
Meanwhile, from the perspective of algorithm acceleration, there is still a possibility that a quantum annealer can be used to generate samples faster than a classical computer. 
Further studies are required to show how much faster it would operate.
If any merit of the quantum AAN exists, the difference stems from the nontrivial quantum device, as in the previous study of investigating the nonequilibrium behavior \cite{Bando2021}.
We will investigate the performance of the quantum AAN via quantum sampling from the D-Wave quantum annealer in comparison with the result obtained via QMC.

There are two directions for further research. The first is to examine applications to other machine learning algorithms, such as reinforcement learning\cite{ayanzadeh2020reinforcement} and Bayesian statistics. The second is to try to apply other quantum sampling methods, such as the sing Born machine\cite{coyle2020born}, to machine learning. 
Such studies will contribute to the development of machine learning with quantum computation.
\section*{Acknowledgements}
The authors would like to thank Manaka Okuyama for useful discussion.
M. O. thanks financial support from the Next Generation High-Performance Computing Infrastructures and Applications R $\&$ D Program by MEXT, and MEXT-Quantum Leap Flagship Program Grant Number JPMXS0120352009.

\newpage
\bibliographystyle{jpsj}
\bibliography{mybibliography.bib}
\end{document}